\documentstyle{mn}
\input{epsf}
\input{rotate}

\title[Velocity dispersions of dwarf spheroidals]{Velocity dispersions of
dwarf spheroidal galaxies: dark matter versus MOND}

\author[Ewa L. {\L}okas]{Ewa L. {\L}okas\\ Nicolaus Copernicus Astronomical
Center, Bartycka 18, 00--716 Warsaw, Poland}

\begin{document}

\maketitle

\begin{abstract}
We present predictions for the line-of-sight velocity dispersion profiles
of dwarf spheroidal galaxies and compare them to observations in the case
of the Fornax dwarf. The predictions are made in the framework of standard
dynamical theory of spherical systems with different velocity
distributions. The stars are assumed to be distributed according to
S\'ersic laws with parameters fitted to observations. We compare
predictions obtained assuming the presence of dark matter haloes (with
density profiles adopted from $N$-body simulations) versus those resulting
from Modified Newtonian Dynamics (MOND). If the anisotropy of velocity
distribution is treated as a free parameter observational data for Fornax 
are reproduced equally well by models with dark matter and with MOND. If stellar 
mass-to-light ratio of $1 M_{\sun}/L_{\sun}$ is assumed,
the required mass of the dark halo is $1.5 \times 10^9
M_{\sun}$, two orders of magnitude bigger than the mass in stars. The
derived MOND acceleration scale is $a_0 = 2.1 \times
10^{-8}$ cm/s$^2$. In both cases a certain amount of tangential anisotropy
in the velocity distribution is needed to reproduce the shape of the
velocity dispersion profile in Fornax.
\end{abstract}


\section{Introduction}

Dwarf spheroidal (dSph) galaxies provide a unique testing field
for the theories of structure formation. Due to their large velocity 
dispersions they are believed to be dominated by dark matter (hereafter DM, 
for a review see Mateo 1997). Because of their small masses they also
lie in the regime of small accelerations, where, according to an
alternative theory of Milgrom (1983a), the Modified Newtonian Dynamics
(MOND) may operate causing a similar increase of velocity dispersions.

Until recently the measurements of velocity dispersion in dwarfs were
restricted to a single value, the central velocity dispersion, on which
all of the dynamical modelling was based. In order to estimate the
mass-to-light ratio in such case a simplifying assumption of isotropic
velocity distribution was usually made. Using this approach many authors
(see Mateo 1998 and references therein)
estimated mass-to-light ratios in dwarfs to be much larger than
stellar values and interpreted this as a signature of the presence of DM. 
However, Milgrom (1995) analysed the data concerning seven well
known dSph  galaxies and concluded that when MOND
is applied their mass-to-light ratios are consistent with stellar values
if all the observational errors are properly taken into account.

Given the velocity dispersion data as a function of distance, which are
becoming available now, we are able to extract much more information from
the observations. Our conclusions do not have to rely on the assumed velocity
anisotropy model, but the best-fitting model can actually be found. On the
theoretical side some progress has also been achieved. The DM
distribution in haloes can now be reliably modelled using the results of
$N$-body simulations (e.g. Navarro, Frenk \& White 1997, hereafter NFW).

The purpose of this work is to explore the two hypotheses, based on DM
and on MOND, by calculating the predicted velocity dispersion profiles
and comparing them to available data. In the case of DM  our analysis is
somewhat similar to an earlier study of Pryor \& Kormendy (1990) who 
also considered two-component models involving stars and DM.
In the case of MOND, our approach here is related to that of Milgrom
(1984) who applied the Jeans equation to study density profiles of
equilibrium systems with isotropic and isothermal velocity
distributions. Here we assume instead the observed density distribution of stars
to obtain the velocity dispersions in MOND for different velocity
anisotropy models.

\vspace{-0.3in}

\section{Velocity dispersion profile}

The radial velocity dispersion $\sigma_{\rm r}(r)$ of stars can be obtained
by solving the Jeans equation (Binney \& Tremaine 1987)
\begin{equation}    \label{m1}
    \frac{\rm d}{{\rm d} r}  (\nu \sigma_{\rm r}^2) + \frac{2 \beta}{r} \nu
	\sigma_{\rm r}^2 = \nu g,
\end{equation}
where $\nu(r)$ is a 3D density of stars, $g(r)$ is the
gravitational acceleration associated with the potential $\Phi(r)$, $g = -
{\rm d} \Phi/{\rm d} r$, and
$\beta=1-\sigma_\theta^2(r)/\sigma_{\rm r}^2(r)$ is a measure of the
anisotropy in the velocity distribution.

In the absence of direct measurements of velocity anisotropy in dSph
 galaxies we have to consider different models for $\beta$.
Our choice here
is guided by the results of $N$-body simulations of DM  haloes
and observations of elliptical galaxies. For DM  haloes
Thomas et al. (1998) find that, in a variety of
cosmological models, the ratio $\sigma_\theta/\sigma_{\rm r}$ is not far
from unity and decreases slowly with distance from the centre to reach
$\simeq 0.8$ at the virial radius. Observations of elliptical galaxies are
also consistent with the orbits being isotropic near the centre and
somewhat radially anisotropic farther away, although cases with tangential
anisotropy are also observed (Gerhard et al. 2001).

Our fiducial anisotropy model
will therefore be that of isotropic orbits: $\sigma_\theta(r)=\sigma_{\rm
r}(r)$ and $\beta=0$. A more realistic approximation is provided by a model
proposed by Osipkov (1979) and Merritt (1985) with $\beta$ dependent on
distance from the centre of the object
\begin{equation} \label{m2}
    \beta_{\rm OM} = \frac{r^2}{r^2 + r_{\rm a}^2} ,
\end{equation}
where $r_{\rm a}$ is the anisotropy radius determining the transition
from isotropic orbits inside to radial orbits outside. This model covers
a wide range of possibilities from isotropic orbits ($r_{\rm a} \rightarrow
\infty$) to radial orbits ($r_{\rm a} \rightarrow 0$). Another possibility
is that of tangential orbits with the fiducial case of circular orbits
when $\beta \rightarrow - \infty$. To cover intermediate cases between the
isotropic and circular orbits we will consider a simple model of
$\beta=$const$<0$.

The solution of the Jeans equation (\ref{m1}) with the boundary condition
$\sigma_{\rm r} \rightarrow 0$ at $r \rightarrow \infty$ for $\beta_{\rm OM}$
anisotropy is
\begin{equation}	\label{m4a}
	\nu \sigma_{\rm r}^2 (\beta=\beta_{\rm OM}) = - \frac{1}{r^2 +r_{\rm a}^2}
	\int_r^\infty (r^2 +r_{\rm a}^2) \nu g \ {\rm d} r ,
\end{equation}
while for $\beta$=const we find
\begin{equation}	\label{m4b}
	\nu \sigma_{\rm r}^2 (\beta={\rm const})= - r^{-2 \beta}
	\int_r^\infty r^{2 \beta} \nu g \ {\rm d}r .
\end{equation}

From the observational point of view, an interesting, measurable quantity
is the line-of-sight velocity dispersion obtained from the 3D velocity
dispersion by integrating along the line of sight (Binney \& Mamon 1982)
\begin{equation}    \label{m3}
    \sigma_{\rm los}^2 (R) = \frac{2}{I(R)} \int_{R}^{\infty}
    \left( 1-\beta \frac{R^2}{r^2} \right) \frac{\nu \,
    \sigma_{\rm r}^2 \,r}{\sqrt{r^2 - R^2}} \,{\rm d} r \ ,
\end{equation}
where $I(R)$ is the surface brightness.
For circular orbits, $\sigma_{\rm r}=0$, and one has
\begin{equation}    \label{m4}
    \sigma_{\rm los}^2 (R) = \frac{1}{I(R)} \int_{R}^{\infty}
    \left( \frac{R}{r} \right)^2 \frac{\nu \,V^2 \,r}{\sqrt{r^2 -
    R^2}}\, {\rm d} r \ ,
\end{equation}
where $V=(r g)^{1/2}$ is the circular velocity associated with
the gravitational acceleration $g(r)$.

Introducing results (\ref{m4a}) or (\ref{m4b}) into equation (\ref{m3})
and inverting the order of integration the calculations of $\sigma_{\rm
los}$ can be reduced to one-dimensional numerical integration of a formula
involving elementary functions in the case of $\beta_{\rm OM}$ (see Mamon \&
\L okas, in preparation) and special functions for arbitrary
$\beta=$const.

We will now discuss different factors entering in the calculations
of $\sigma_{\rm los}$, namely the density distribution of stars $\nu$, the
mass distributions contributed by stars and DM  to gravitational
acceleration $g$ and how this acceleration is changed in MOND.

\vspace{-0.1in}

\subsection{Stars}

The distribution of stars is well known in a number of dSph galaxies
from surface brightness measurements and is usually well fitted by the
exponential law or its generalization i.e. the S\'ersic profile
(S\'ersic 1968, see also Ciotti 1991)
\begin{equation}    \label{m5}
	I(R) = I_0 \exp [-(R/R_{\rm S})^{1/m}],
\end{equation}
where $I_0$ is the central surface brightness and $R_{\rm S}$ is the
characteristic projected radius of the S\'ersic profile. The S\'ersic
profile with parameter $m$ varying in the range $1 \le m \le 10$ has been
found (Caon, Capaccioli \& D'Onofrio 1993) to fit surface brightness
profiles of elliptical galaxies in large mass range better than the
standard de Vaucouleurs law (corresponding to $m=4$). For dSph
systems the fits are usually performed with $m=1$, although is some
cases other values of $m$ are found to provide better fits
(e.g. Caldwell 1999).

The 3D luminosity density $\nu(r)$ is obtained from $I(R)$ by
deprojection
\begin{equation}	\label{m5a}
	\nu(r) = - \frac{1}{\pi} \int_r^\infty \frac{{\rm d} I}{{\rm d} R}
	\frac{{\rm d} R}{\sqrt{R^2 - r^2}}.
\end{equation}
In the case of $m=1$ we get $\nu(r, m=1) = I_0 K_0 (r/R_{\rm S})/(\pi R_{\rm S})$,
where $K_0(x)$ is the modified Bessel function of the second kind. For
other values of $m$ in the range $1/2 \le m \le 10$ an excellent
approximation for $\nu(r)$ is provided by (Lima Neto, Gerbal \&
M\'arquez 1999)
\begin{eqnarray}
	\nu(r) &=& \nu_0 \left( \frac{r}{R_{\rm S}} \right)^{-p}
	\exp \left[-\left( \frac{r}{R_{\rm S}}
	\right)^{1/m} \right] \label{m5c} \\
	\nu_0 &=& \frac{I_0 \Gamma(2 m)}{2 R_{\rm S}
	\Gamma[(3-p) m]}  \nonumber \\
	p &=& 1.0 - 0.6097/m + 0.05463/m^2 \nonumber.
\end{eqnarray}

The mass distribution of stars following from (\ref{m5c}) is
\begin{equation}	\label{m5d}
	M_{*}(r) = \Upsilon L_{\rm tot} \frac{\gamma[(3-p)m,
	(r/R_{\rm S})^{1/m}]}{\Gamma[(3-p)m]},
\end{equation}
where $\Upsilon$=const is the mass-to-light ratio for stars, $L_{\rm tot}$
is the total luminosity of the galaxy and $\gamma(\alpha, x) = \int_0^x
{\rm e}^{-t} t^{\alpha-1} {\rm d} t$ is the incomplete gamma function.

\vspace{-0.1in}

\subsection{Dark matter}

We assume that the density distribution of DM
 is given by the so-called universal profile (NFW, see also \L okas \& Mamon 2001)
\begin{equation}    \label{m6}
    \frac{\rho(r)}{\rho_{\rm crit,0}} = \frac{\delta_{\rm char}}{(r/r_{\rm
    s})\,(1+r/r_{\rm s})^{2}}
\end{equation}
with a single fitting parameter $\delta_{\rm char}$, the characteristic
density. The scale radius $r_{\rm s}$ is defined by $r_{\rm s} =
r_{\rm v}/c$ where $r_{\rm v}$ is the virial radius (the distance from the centre
of the halo within which the mean density is
$200$ times the present critical density, $\rho_{\rm crit,0}$) and $c$ is the
concentration parameter related to the
characteristic density by $\delta_{\rm char} = 200 c^3 g(c)/3$
with $g(c) = 1/[\ln (1+c) - c/(1+c)]$. The value of $c$ is known to
depend on mass and will be extrapolated here to small masses
characteristic of dwarf galaxies assuming the following formula fitted to
$N$-body results for $\Lambda$CDM cosmology of Jing \& Suto (2000)
\begin{equation}	\label{m7}
	c = 10.23 \left( \frac{h M_{\rm v}}{10^{12} M_{\sun}} \right)^{-0.088},
\end{equation}
where $M_{\rm v} = 800 \pi r_{\rm v}^3 \rho_{\rm crit,0}/3$ is
the mass within virial radius
and $h$ is the Hubble constant in units of 100 km s$^{-1}$ Mpc$^{-1}$.
In the following we will use $h=0.7$.

The mass distribution associated with the NFW profile (\ref{m6}) is
\begin{equation}	\label{m8}
	M_{\rm NFW}(r) = g(c) M_{\rm v} \left[ \ln (1 + c s) -
	\frac{c s}{1 + c s} \right],
\end{equation}
where $s=r/r_{\rm v}$ and for $h=0.7$ the virial radius scales with the virial
mass so that $r_{\rm v}=206 [ M_{\rm v}/(10^{12} M_{\sun})]^{1/3} $ kpc.

\vspace{-0.1in}

\subsection{MOND}

In the case of MOND the gravitational acceleration is changed so that
(Milgrom 1984)
\begin{equation}	\label{m10}
   g_{\rm M} = \frac{g_{\rm N}}{\mu(g_{\rm N}/a_0)},
\end{equation}
where $g_{\rm M}$ and $g_{\rm N}$ are the modified and Newtonian gravitational
accelerations, respectively. $a_0$ is the characteristic acceleration
scale of MOND estimated to be of the order of $10^{-8}$ cm/s$^2$. The
function $\mu(x)$ has to reproduce Newtonian dynamics at large
accelerations, i.e. $\mu(x) \rightarrow 1$ for $x \rightarrow \infty$, and
is supposed to be $\mu(x) \approx x$ for $x \rightarrow 0$. A good
approximation is then provided by e.g. $\mu(x)=x(1+x^2)^{-1/2}$ (Milgrom
1983b). Using this approximation we find from equation (\ref{m10})
\begin{equation}	\label{m11}
	g_{\rm M} = \left[\frac{g_{\rm N}^2 + g_{\rm N}
	(4 a_0^2 + g_{\rm N}^2)^{1/2}}{2} \right]^{1/2}.
\end{equation}

\vspace{-0.2in}

\section{Results for the Fornax dwarf}

The structural parameters of the Fornax dwarf needed in the following
computation were taken from Irwin \& Hatzidimitriou (1995, hereafter IH).
They fitted the surface brightness profile of Fornax with the exponential
distribution
($m=1$ in equation [\ref{m5}]) finding the S\'ersic radius ($r_{\rm e}$ in their
notation) $R_{\rm S}=9.9$ arcmin $=0.35$ kpc assuming the distance modulus
$m-M=20.4$. The brightness of Fornax in V-band is $M_{\rm V} = -13.0$ mag so the
total luminosity needed in equation (\ref{m5d}) is $L_{\rm tot,V} = 1.4 \times 10^7
L_{\sun}$. We also adopt the mass-to-light ratio for stars in this band to
be $\Upsilon_{\rm V} \approx 1 M_{\sun}/L_{\sun}$ (Mateo et al. 1991).

\begin{figure}
\begin{center}
    \leavevmode
    \epsfxsize=7.8cm
    \epsfbox[50 50 320 560]{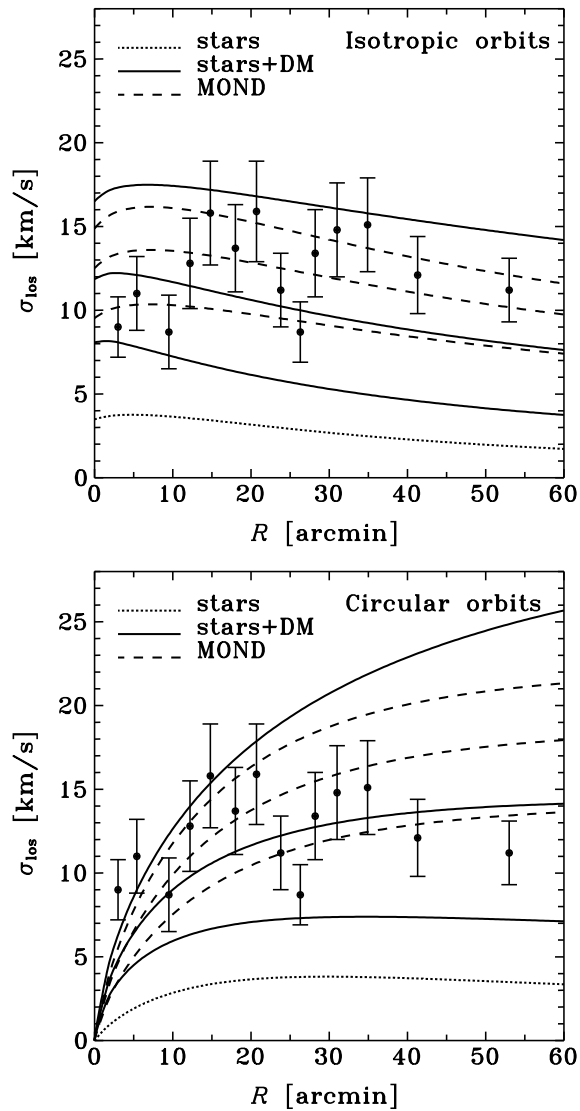}
\end{center}
\caption{Upper panel: predicted line-of-sight velocity dispersion profiles
for the Fornax dwarf in the case of isotropic orbits. The dotted line
shows the prediction obtained by solving Jeans equation for stars only.
The three solid lines correspond to stars + DM  with the dark halo
masses of $10^8$, $10^9$ and $10^{10}$ $M_{\sun}$ from bottom to top. The
three dashed lines show results for stars in MOND with $a_0$ = 1, 3, 6
$\times 10^{-8}$ cm/s$^2$ from bottom to top. The data for Fornax are from
Mateo (1997). Lower panel: same as upper panel, but for circular orbits.}
\label{fornax}
\vspace{-0.1in}
\end{figure}

Figure~\ref{fornax} shows the predicted line-of-sight velocity dispersion
profiles for the Fornax dwarf together with the data from Mateo (1997).
The two panels present results for different velocity anisotropy models.
The upper panel gives the results in the case of isotropic orbits obtained
from equations (\ref{m4a}) and (\ref{m3}) in the limit of large $r_{\rm a}$. The
lower panel is for circular orbits showing results from
(\ref{m4}). The three types of curves in each panel correspond to the
three possibilities of the velocity dispersion profile being generated
1) by stars moving under Newtonian gravitational acceleration generated by
stars only (dotted curves), 2) by stars moving under Newtonian
gravitational acceleration generated by stars and DM  (solid
lines), 3) by stars moving under modified (MONDian) acceleration generated
by stars (dashed lines).

For all cases the 3D distribution of stars $\nu(r)$ is given by
(\ref{m5c}). The gravitational acceleration for the Newtonian cases 1) and
2) is $g=g_{\rm N}=-G M(r)/r^2$ with $M(r)=M_*(r)$ for 1) and $M(r)=M_*(r) +
M_{\rm NFW}(r)$ for 2), where $M_*(r)$ and $M_{\rm NFW}(r)$
are given by equations
(\ref{m5d}) and (\ref{m8}), respectively. In the case of MOND we have
$g=g_{\rm M}$ from equation (\ref{m11}) with $g_{\rm N}=-G M_* (r)/r^2$.
The three
solid curves in each panel correspond from bottom to top to the virial
masses of the dark halo $M_{\rm v} = 10^8$, $10^9$ and $10^{10}$ $M_{\sun}$.
The three dashed lines show from bottom to top results obtained with different
characteristic MOND acceleration scales $a_0$ = 1, 3, 6 $\times 10^{-8}$
cm/s$^2$.

\begin{figure}
\begin{center}
    \leavevmode
    \epsfxsize=8cm
    \epsfbox[50 50 320 310]{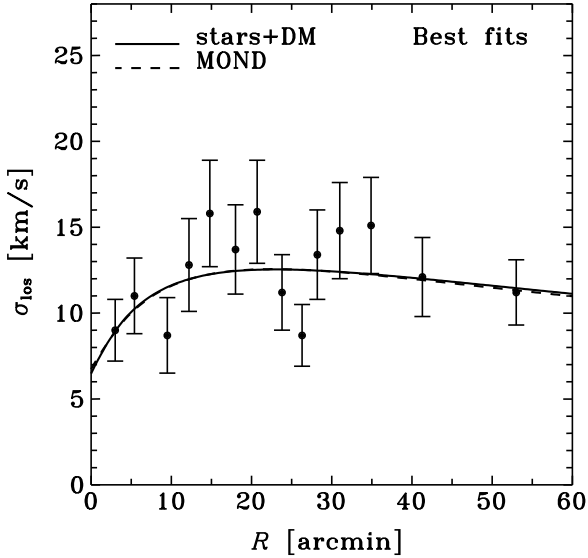}
\end{center}
\caption{Best-fitting models for Fornax based on stars+DM (solid line) and
MOND (dashed line). The model involving DM  has $\beta=-1.4$ and
$M_{\rm v}=1.5 \times 10^9 M_{\sun}$. The model with MOND requires $\beta=-1$
and $a_0 = 2.1 \times 10^{-8}$ cm/s$^2$.}
\label{fornaxtw}
\vspace{-0.1in}
\end{figure}

Figure~\ref{fornax} clearly shows that the stars alone cannot reproduce
the observed velocity dispersion profile of Fornax, unless the stellar
mass-to-light ratio is much larger than assumed. In the following we will
farther analyse the cases 2) and 3) trying to find the best-fitting
models. Comparing the predictions shown in Figure~\ref{fornax} with the data
we find that a certain amount of tangential anisotropy is needed to
reproduce the shape of the observed velocity dispersion profile. Indeed,
trying more radial anisotropy than in the case of isotropic orbits by
using small values of $r_{\rm a}$ (comparable to $R_{\rm S}$ or $r_{\rm v}$) in
(\ref{m4a}) and (\ref{m3}) makes the curves in the upper panel of
Figure~\ref{fornax} decrease even more steeply with distance, contrary to the
trend observed in the data. Therefore in what follows we adopt the
velocity anisotropy model of $\beta=$const$ < 0$ and treat $\beta$ as a
free parameter.

To find the best models we performed a two-parameter fitting for each
case. The fitting parameters were $(M_{\rm v}, \beta)$ for stars+DM
case and $(a_0, \beta)$ for MOND. The results of minimizing $\chi^2$
weighted by the variances of the data points are shown in
Figure~\ref{fornaxtw}. It turns out that the DM  model fits the
data best for  $\beta=-1.4$ and $M_{\rm v}=1.5 \times 10^9 M_{\sun}$. This
virial mass is more than two orders of magnitude larger than the
stellar mass ($1.4 \times 10^7 M_{\sun}$ for $\Upsilon_{\rm V} =
1 M_{\sun}/L_{\sun}$). In the case of MOND the best-fitting parameters are
$\beta=-1$ and $a_0 = 2.1 \times 10^{-8}$ cm/s$^2$. This value of $a_0$ is
consistent with other estimates (Milgrom 1983b).

\begin{figure}
\begin{center}
    \leavevmode
    \epsfxsize=8cm
    \epsfbox[50 50 320 310]{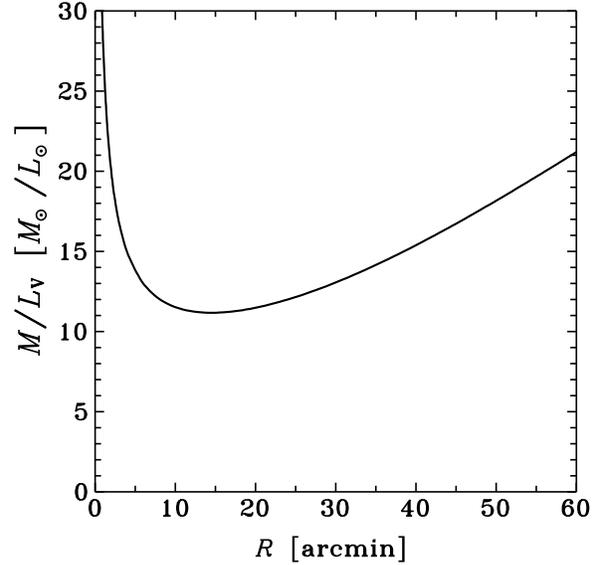}
\end{center}
\caption{The dependence of the mass-to-light ratio of Fornax
on distance. The mass comes from stars and DM  with the best fitting
$M_{\rm v}=1.5 \times 10^9 M_{\sun}$.}
\label{mtl}
\vspace{-0.1in}
\end{figure}

Figure~\ref{mtl} shows the combined mass-to-light ratio in V-band
for Fornax calculated from
\begin{equation}	\label{m12}
	M/L_{\rm V} = \frac{M_{\rm NFW}(r) + M_* (r)}{L_{\rm V}(r)},
\end{equation}
where $M_{\rm NFW}(r)$ and $M_* (r)$ are given
by equations (\ref{m8}) and (\ref{m5d})
respectively, and $L_{\rm V}(r) = M_* (r)/\Upsilon_{\rm V}$
is the luminosity distribution. The
shape of the curve reflects the fact that DM  dominates both at small
and large distances. The divergence of $M/L$ at the centre of the galaxy is due
to the fact that the luminosity
density (\ref{m5c}) increases towards the centre more slowly than the
DM  cusp of $r^{-1}$ in (\ref{m6}). The growth of $M/L$ at large
distances is due to the behaviour of $M_{\rm NFW} (r)$ which diverges logarithmically
while the mass in stars is finite. The minimum of $M/L_{\rm V} \approx 11$
appears at a scale of the order of the S\'ersic radius and is consistent with
the values of the central mass-to-light ratio estimated by Mateo et al. (1991).

\vspace{-0.2in}

\section{Discussion}

We have found that the predictions based on MOND with $a_0 = 2.1 \times
10^{-8}$ cm/s$^2$ fit the Fornax data equally well as models based on
DM  with virial mass of $M_{\rm v}=1.5 \times 10^9 M_{\sun}$.
In both cases a certain amount of tangential anisotropy is
needed to reproduce the shape of the observed velocity dispersion profile.
It turned out that the MOND predictions in the case of Fornax dwarf need
less tangential anisotropy than DM  predictions to reproduce
the shape of the velocity dispersion profile. It remains to be checked
how the DM predictions depend on the particular shape of DM distribution.

The particular fitted values of $a_0$ and $M_{\rm v}$ are of course
sensitive to the observational parameters assumed. The measured
luminosity, distance and structural parameters of the dwarfs are all known
with limited accuracy, but the detailed analysis of the errors in them is
beyond the scope of this paper. In the case of MOND the
most doubtful input parameter is the stellar mass-to-light ratio in V-band.
We have assumed here $\Upsilon_{\rm V} \approx 1 M_{\sun}/L_{\sun}$ 
characteristic for a relatively young stellar population of Fornax, 
however this value may well
be different by a factor of two (Mateo et al. 1991).
To estimate its effect we performed a similar MOND fitting
as described in the previous section with $\Upsilon_{\rm V} = 0.5$ and $2 \
M_{\sun}/L_{\sun}$, obtaining respectively $a_0 = 4.3 $ and $1.1 \ \times
10^{-8}$ cm/s$^2$ with the same best-fitting $\beta$.
If $\Upsilon_{\rm V} = 0.5 M_{\sun}/L_{\sun}$ is indeed the
lower limit of mass-to-light ratio, then $a_0 = 4.3 \times 10^{-8}$
cm/s$^2$ is the upper limit for $a_0$.

In the case of DM  the stellar mass-to-light is not likely to
influence much the results since the input from stars to the velocity
dispersion is small. Here the most doubtful parameter is the
concentration of DM  haloes, which we adopted to follow the same
dependence on mass as larger haloes actually studied in $N$-body
simulations. Given, however, that this dependence is weak in the
$\Lambda$CDM Universe, the possible error associated with this factor is
rather small.

One more reason to worry are the tidal interactions that may affect the
dynamics of some dwarfs and our interpretations of their velocity
dispersions. Piatek \& Pryor (1995) have performed numerical simulations
of such effects and concluded that in Newtonian dynamics they
should not affect much the inferred mass-to-light ratios (but see Kroupa 1997). 
However, Brada \& Milgrom (2000) found that in MOND, due to its non-linearity,
a galaxy system can be more affected by an external field.

\begin{figure}
\begin{center}
    \leavevmode
    \epsfxsize=8.5cm
    \epsfbox[50 50 540 310]{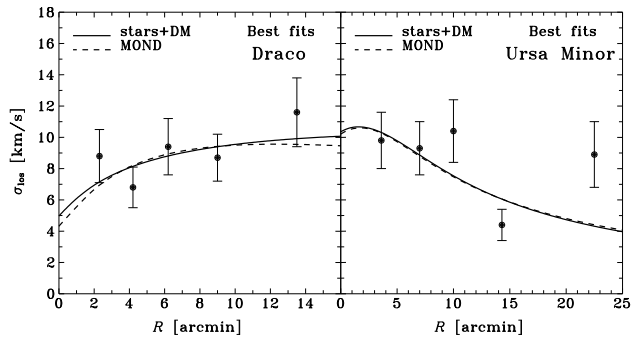}
\end{center}
\caption{Best-fitting models for Draco (left panel) and Ursa Minor (right panel)
based on stars+DM (solid lines) and MOND (dashed lines). The data are from
Armandroff et al. (1997)}
\label{du}
\vspace{-0.1in}
\end{figure}

One may ask if similar results hold for other dSph galaxies as well.
Of those studied by IH the velocity dispersion profiles with a few data 
points are available only for Sextans, Ursa Minor and
Draco dwarfs (Hargreaves et al. 1994a, 1994b, 1996; Armandroff, Olszewski \& Pryor 
1995; Armandroff, Pryor \& Olszewski 1997). Applying the same analysis as we did for
Fornax to the 5 data points for Draco and Ursa Minor from Armandroff et al. (1997)
(with $\Upsilon_{\rm V} =1 M_{\sun}/L_{\sun}$) we find that again the MOND
and DM  models can fit the data equally well. For Draco the best-fitting models 
have tangential anisotropy, we get $\beta= -0.9$, $M_{\rm v}=4.2 \times 10^9 M_{\sun}$ 
($2 \times 10^4$ the mass in stars) for DM and $\beta=-1.5$, $a_0 = 50 \times 
10^{-8}$ cm/s$^2$ for MOND. For Ursa Minor the best fits require a certain
amount of {\em radial\/} anisotropy which we model with (\ref{m2}). 
We get $r_{\rm a} = 0.014 r_{\rm v} = 0.86 R_{\rm S} = 0.16$ kpc,
$M_{\rm v}=1.9 \times 10^8 M_{\sun}$ ($10^3$ the mass in stars)
for DM  and $r_{\rm a} =0.012 r_{\rm v} = 0.76 R_{\rm S} = 0.14$
kpc, $a_0 = 12 \times 10^{-8}$ cm/s$^2$ for MOND. The predictions of the best-fitting
models together with the data are shown in Figure~\ref{du}.

The case of Ursa Minor is doubtful since it has been recently observed to possess
tidal tails (Mart\'{\i}nez-Delgado et al. 2001) and therefore may not be in virial 
equilibrium. The high value of $a_0$ estimated for Draco can, however, pose a 
serious problem for MOND (we agree on this point with Gerhard \& Spergel 1992)
since $a_0$ is supposed to be a universal constant. Bringing 
the value down by an order of magnitude would require very high stellar mass-to-light 
ratio $\Upsilon_{\rm V} \sim 10 M_{\sun}/L_{\sun}$ or we would have to believe that the
measurements seriously overestimate the velocity dispersion.

\vspace{-0.2in}

\section*{Acknowledgements}

I wish to thank A. Graham, P. Kroupa, G. Mamon, C. Pryor, P. Salucci and
the referee, M. Mateo, for useful comments and suggestions which helped to improve
the paper.
I am also grateful to Y. P. Jing for providing the results for the
concentration of DM  haloes in numerical form. This research was
partially supported by the Polish State Committee for Scientific Research
grant No. 2P03D02319.

\end{document}